\documentclass[aps,prb,twocolumn,superscriptaddress,floatfix]{revtex4}
\usepackage{graphicx,amsmath,amssymb}
\def\abs#1{\vert#1\vert}

\def\Cone{C_2}
\def\Ctwo{C_3}
\def\Cthree{C_1}
\begin{document}

\title{Correlation functions and excitation spectrum of the
frustrated ferromagnetic spin-$\frac{1}{2}$ chain in an external magnetic field}

\author{T. Vekua}
\affiliation{Laboratoire de Physique Th\'eorique et Mod\`eles Statistiques, Universit\'e Paris Sud, 91405 Orsay Cedex, France}
  
\author{A. Honecker}
\affiliation{Institut f\"ur Theoretische Physik, Universit\"at G\"ottingen,
37077 G\"ottingen, Germany}

\author{H.-J. Mikeska}
\affiliation{Institut f\"{u}r Theoretische Physik,
Universit\"{a}t Hannover, Appelstrasse 2, 30167 Hannover, Germany}

\author{F. Heidrich-Meisner}
\affiliation{Materials  Science and Technology Division, Oak Ridge National Laboratory,
  Tennessee, 37831, USA and\\
 Department of Physics and Astronomy, University of Tennessee, Knoxville,
 Tennessee 37996, USA}

\date{April 5, 2007; revised: July 6, 2007}

\begin{abstract}
Magnetic field effects on the one-dimensional frustrated
ferromagnetic chain are studied by means of effective field theory
approaches in combination with numerical calculations utilizing Lanczos
diagonalization and the density matrix renormalization group method. The
nature of the ground state is shown to change from a spin-density-wave
region to a nematic-like one upon approaching the saturation
magnetization. The excitation spectrum is analyzed and the behavior of the single
spin-flip excitation gap is studied in detail, including the emergent finite-size corrections.

\end{abstract}

\maketitle

\section{Introduction}

The interest in helical and chiral phases of frustrated low-dimensional 
quantum magnets has
been triggered by recent experimental results. While many copper-oxide based materials predominantly realize
antiferromagnetic exchange interactions,  
several candidate materials with magnetic
properties believed to be described by frustrated ferromagnetic
chains have been identified,\cite{Hase04,Gibson04,Enderle05,Banks06,buettgen07,Drechsler07}
including  Rb$_2$Cu$_2$Mo$_3$O$_{12}$ (Ref.~\onlinecite{Hase04}),  LiCuVO$_4$ 
(Refs.~\onlinecite{Gibson04,Enderle05,Banks06,buettgen07}), and Li$_2$ZrCuO$_4$ (Ref.~\onlinecite{Drechsler07}).
The frustrated antiferromagnetic chain is well-studied,\cite{review-frust-1D} but the magnetic phase diagram
of the model with ferromagnetic nearest-neighbor interactions remains a subject of 
active theoretical investigations.\cite{Chubukov91,Heidrich-Meisner06,Kuzian,Dmitriev}

In this work  we consider a parameter regime that is  in particular relevant for  the low-energy properties of
LiCuVO$_4$, corresponding to a ratio of $J_1\approx -0.3\,J_2$ between the
nearest neighbor interaction $J_1$ and the frustrating next-nearest neighbor interaction $J_2>0$. As
the interchain couplings for this material are an order of magnitude smaller
than the intrachain ones,\cite{Enderle05} we analyze a purely one-dimensional (1D) model.
Apart  from mean-field based predictions, \cite{Chubukov91} the
nature of the ground state in a magnetic field $h$ is not yet completely known.
Therefore, combining the bosonization technique with a numerical analysis we determine
ground-state properties and discuss the model's  elementary excitations.

The Hamiltonian for our 1D model reads:
\begin{equation}
H= \sum_{x=1}^L \left( J_1 \vec{S}_x \cdot \vec{S}_{x+1}
 + J_2 \vec{S}_x \cdot \vec{S}_{x+2} \right) -h \sum_x S_x^z \, ,
\label{eq:1.1}
\end{equation}
where  $\vec{S}_x$ represents a spin one-half operator at site $x$.

Bosonization has turned out to be 
the appropriate language for describing the regime $|J_1|\ll J_2$
of Eq.~(\ref{eq:1.1}). This result has been established
by studying the magnetization process yielding a good agreement
between field theory and numerical data.\cite{Heidrich-Meisner06}
The derivation of the effective field theory is summarized in Sec.~\ref{sec:field}.
Here, we extend on such comparison of analytical and numerical results and further
confirm the predictions of field theory by analyzing several correlation
functions in Sec.~\ref{sec:corrs}. Then, in Sec.~\ref{sec:ex}, we numerically compute the one- and two-spin flip excitation
gaps and compare them to field-theory predictions. Finally, Sec.~\ref{sec:sum} contains a summary and a discussion of our results.

\section{Effective field theory}
\label{sec:field}
We start from an effective field theory describing the long-wavelength
fluctuations of Eq.~(\ref{eq:1.1}). In the limit of strong next-nearest
neighbor interactions $J_2 \gg \abs{J_1}$, the spin operators can
be expressed as:
\begin{eqnarray}
\label{BosFor}
S^z_{\alpha}(r)& \sim&m+c(m)\sin \big\{2k_F r+ \sqrt{4\pi } \phi_{\alpha}\big\}+\cdots \nonumber\\
S_{\alpha}^-(r)& \sim& (-1)^r {\rm e}^{-i\theta_{\alpha}\sqrt{\pi}}+\cdots\,.
\end{eqnarray}
$k_F=(\frac{1}{2}-m)\pi$ is the Fermi-wave vector and $\alpha=1,2$ 
enumerates the two chains of the zig-zag ladder.
In relation with Eq.~(\ref{eq:1.1}), note that
$\vec{S}_{1}(r)=\vec{S}_{(x+1)/2}$ ($\vec{S}_{2}(r)=\vec{S}_{x/2}$)
for $x$ odd (even).
$\phi_{\alpha}$ and
$\theta_{\alpha}$ are compactified quantum fields describing the out-of-plane
and in-plane angles of fluctuating spins obeying Gaussian Hamiltonians:
\begin{equation}
\label{SpinChainBosHam}
{\cal H} =  \frac{v}{2}\int dx \, \Big\{\frac{1}{K}(\partial_x \phi_{\alpha})^{2} 
+ K (\partial_x \theta_{\alpha})^{2}\Big\} \, ,
\end{equation}
with $[\phi_{\alpha}(x),\theta_{\alpha}(y)] = i\Theta (y-x)$,
where $\Theta(x)$ is the Heaviside function. Sub-leading terms  are suppressed in Eq.~(\ref{BosFor}). $m$ is the
magnetization of decoupled chains, related to the real magnetization $M$
of the zig-zag system by:
\begin{equation}
\label{Mag}
M\simeq m\left(1-\frac{2K(m)J_1}{\pi v(m)}\right) \, .
\end{equation}
$K(m)$ and $v(m)$ are the Luttinger liquid (LL) parameter and the spin-wave
velocity of the decoupled chains, respectively.
The nonuniversal amplitude $c(m)$ appearing in 
the bosonization formulas (\ref{BosFor}) has been determined
from density matrix renormalization
group (DMRG) calculations.\cite{Hikihara04}
Note that in our notation $M=1/2$ at saturation.

Now we perturbatively add the interchain
coupling term to two decoupled chains, each of which  is described
by an effective Hamiltonian of the form Eq.~(\ref{SpinChainBosHam})
and fields $\phi_i$ and $\theta_i$, $i=1,2$. 
For convenience, we transform to the symmetric and antisymmetric combinations of the bosonic fields
$\phi_{\pm}=(\phi_1\pm \phi_2)/\sqrt{2}$ and  
$\theta_{\pm}= (\theta_1 \pm \theta_2)/\sqrt{2}$. 
In this basis and apart from terms ${\mathcal H}_{0}^{\pm}$ of the form (\ref{SpinChainBosHam}), the effective Hamiltonian 
describing low-energy properties of Eq.~(\ref{eq:1.1})
contains a single relevant interaction term with the bare coupling $g_1\propto J_1\ll v$:
\begin{eqnarray}
{\mathcal H}_{\rm eff} =
 {\mathcal H}_{0}^{+} +{\mathcal H}_{0}^{-} + g_1\int dx\,\cos\big(k_F+\sqrt{8\pi }\phi_{-}\big)\, ,
\label{symantisym}
\end{eqnarray}
and the renormalized LL parameters $K_{\pm}$ are, in the weak coupling limit:
\begin{equation}
\label{LL}
K_{\pm}=K \, \left( 1 \mp J_1 \, \frac{K}{\pi \, v} \right) \, .
\end{equation}
$K_+$ is the Luttinger-liquid parameter of the soft mode of the zig-zag
ladder.
The Hamiltonian
(\ref{symantisym}) represents the minimal effective low-energy field theory 
describing the region $J_2\gg \abs{J_1} $ of the frustrated FM spin-${1}/{2}$
chain for $M \ne 0$.\cite{Heidrich-Meisner06,Cabra98}
The relevant interaction term $\cos \sqrt{8\pi}\phi_{-}$ opens
 a gap in the $\phi_-$ sector.
Since $S_{x+1}^z-S_{x}^z\sim\partial_x\phi_-$, relative fluctuations of the two chains are locked.
This implies that single-spin flips are gapped  with a
sine-Gordon gap in the sector describing relative spin fluctuations of the
two-chain system.\cite{Heidrich-Meisner06} Gapless excitations come from the
$\Delta S^z=2$ channel, {\it i.e.}\ only those excitations are soft where
spins simultaneously flip on both chains. DMRG results show that this picture applies to a large part of the magnetic phase diagram.\cite{Heidrich-Meisner06}

\section{Correlation functions}
\label{sec:corrs}

We now turn to the ground state properties of Eq.~(\ref{eq:1.1}) as a function of
magnetization, concentrating on several correlation functions in order to identify
the leading instabilities. Note that our analysis is only valid  if $M \neq 0$.
Apart from a term representing the magnetization $M$ 
induced by the external
field, the longitudinal correlation function 
shows an algebraic decay with distance $r$:
\begin{equation}
\label{zzcor}
\langle S_{\alpha}^{z}(0)S_{\beta}^{z}(r) \rangle \simeq
M^2+\frac{\Cthree \cos({2k_Fr+(\alpha-\beta)k_F})}{2\pi^2 r^{K_+}}-\frac{K_+}{8\pi^2 r^2}\,.
\end{equation}
The constants $C_i$, $i=1,2,3$, appearing here and in Eq.~(\ref{nematic}) will be determined through a comparison
with numerical results.

In contrast to Eq.~(\ref{zzcor}), the transverse $xy$-correlation functions
decay exponentially reflecting the gapped nature of the single spin-flip
excitations.  Here we do not restrict ourselves to the equal-time expression only,
because we will need non-equal time correlation functions to extract the finite-size corrections to the gap later on.
We obtain: 
\begin{equation}
\langle S_{\alpha}^{+}(0,0)S_{\beta}^{-}(r,\tau) \rangle \simeq
\frac {\delta_{\alpha,\beta}(-1)^r {\rm e}^{-\Delta_1(M)\sqrt{\tau^2+r^2/v_-^2}}}{(r^2+v_+^2\tau^2)^{\frac{1}{8K_+}}(r^2+v_-^2\tau^2)^{\frac{1}{8K_-}} }
\,,
\label{eq:ssTrans}
\end{equation}
where $\tau$ stands for the Euclidean time, $\Delta_1(M)$ is the $\Delta S^z=1$ gap, and $v_{\pm}\sim  v \pm J_1 /\pi$ in the weak coupling limit.
The Kronecker delta strictly applies to the thermodynamic limit, while
on the lattice an additional contribution for $\alpha \ne \beta$ exists.

It is noteworthy that, different from Eq.~(\ref{eq:ssTrans}), the
in-plane correlation functions involving bilinear spin combinations decay
algebraically. This stems from the gapless nature of $\Delta S^z=2$ excitations. In fact, these are the slowest decaying
correlators close to the saturation magnetization:
\begin{eqnarray}
\label{nematic}
 \langle S_1^+(r) S_2^+(r) \, S_1^-(0) S_2^-(0) \rangle
\simeq\frac{\Cone}{ r^{1/K_+}}+\frac{\Ctwo\cos(2k_Fr)}{r^{K_+ +{1}/{K_+}}} \, .\label{eq:nem}
\end{eqnarray}
This result is reminiscent of a partially ordered state because
the ordering tendencies in this correlation function are more
pronounced than those of the corresponding single-spin correlation
function Eq.~(\ref{eq:ssTrans}). Therefore, we call
the correlator (\ref{nematic}) `nematic'. Furthermore, we will refer
to a situation where Eq.~(\ref{nematic}) is the slowest decaying one
among {\em all} correlation functions as a `nematic-like phase'.

By virtue of the exponential decay in (\ref{eq:ssTrans}),
the correlator (\ref{nematic}) is proportional to:
\begin{equation}
\langle (S_1^+(r) + S_2^+(r))^2 \, (S_1^-(0) + S_2^-(0))^2 \rangle  \,.
\end{equation}
The term $(S_1^{\alpha} + S_2^{\alpha})^2$ appearing in the case of  the $S=\frac{1}{2}$ zig-zag ladder
corresponds to the operator $(S^{\alpha})^2$ in the case of a $S=1$ chain. One can think of an 
effective $S=1$ spin formed from two neighboring $S=\frac{1}{2}$ spins 
coupled by the ferromagnetic interaction. A similar behavior of correlation functions, namely the exponential decay of
in-plane spin components and the algebraic decay of their bilinear combinations,
is encountered also in the  $XY2$ phase of the anisotropic $S=1$ chain\cite{Schulz1} and in the spin-1 chain with biquadratic interactions, see, e.g., Ref.~\onlinecite{LST06}.

The algebraic decay of the nematic correlator as opposed to the exponential
decay of (\ref{eq:ssTrans}) suggests that
there are tendencies towards nematic ordering in this phase.
Depending on the value of $K_+$ the dominant instabilities are either
spin-density-wave ones for $K_+<1$ or nematic ones for $K_+>1$.
{}From the result for $K_+$ given in Eq.~(\ref{LL}) one can perturbatively evaluate the crossover value of $J_1$:
\begin{eqnarray}
\label{crossover}
|J_{1,cr}|= \frac{\pi v(m)}{K(m)}\left(\frac{1}{K(m)}-1\right) \,.
\end{eqnarray}
For $J_1<J_{1,cr}$ the nematic correlator (\ref{nematic}) is the slowest
decaying one, {\it i.e.}\ one is in the nematic-like phase.
The behavior of the cross-over line can be read off from the
behavior of $K(m)$: $K(m)$ increases monotonically with $m$, tends to $K=1$
for $m \to 1/2$, and satisfies $K<1$ for $m<1/2$ (see, e.g.,
Refs.~\onlinecite{Totsuka97,CHP98}).
Therefore, we have $J_{1,cr}=0$ for $M=1/2$ with increasing ferromagnetic
$|J_{1,cr}|$ for decreasing $M$. This means that for $J_1 < 0$ a regime
opens at high $M$ where nematic correlations given by Eq.~(\ref{nematic})
dominate over spin-density-wave correlations given by Eq.~(\ref{zzcor}),
in agreement with Chubukov's prediction.\cite{Chubukov91}

\begin{figure}[t!]
\centerline{\includegraphics[width=\columnwidth]{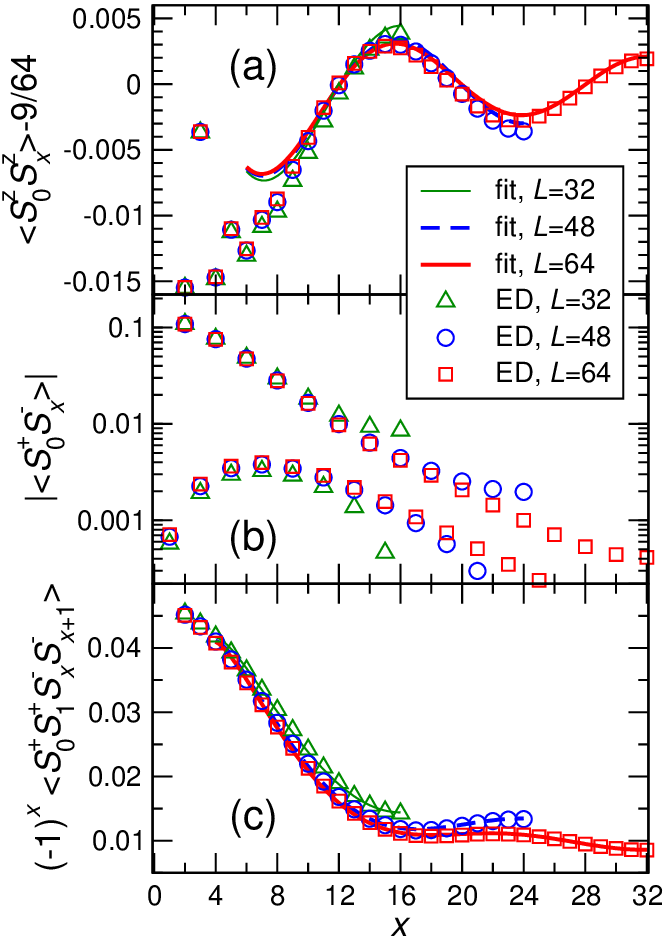}}
\caption{(Color online)
Correlation functions at $J_1 = -J_2 <0 $,  and magnetization
$M=3/8$:
(a) longitudinal component $S^z_x$, (b) transverse component $S^\pm_x$,
(c) spin nematic $S^\pm_x S^\pm_{x+1}$. $x$ is the distance in a
single-chain notation.
ED  results for periodic boundary conditions are shown by
symbols, fits by lines.
Note the logarithmic scale of the vertical axis in panel (b).
}
\label{fig:Correl}
\end{figure}
Now we  check the correlation functions obtained within bosonization
against exact diagonalization (ED) results. Numerical data obtained for $J_1 =
-J_2 <0 $ and $M=3/8$ on finite systems with periodic boundary conditions are
shown in Fig.~\ref{fig:Correl}. This parameter set allows for a clear test of the above predictions, but represents the
generic behavior in the phase of two weakly coupled chains.
To take into account finite-size effects we
use the observation that for a conformally invariant theory, any power law on a
plane becomes a power law in the following variable defined on a cylinder
of circumference $L$:
\begin{equation}
x \to \frac{L}{\pi} \, \sin\left(\frac{x \, \pi}{L}\right) \,.
\label{eq:ConfMap}
\end{equation}

First we fit the nematic correlator given by Eq.~(\ref{nematic}),
which from bosonization is expected to be the leading instability
at high magnetizations.
Using the part with $x \ge 5$ of the $L=64$ data shown in
Fig.~\ref{fig:Correl}c, we find
$1/K_+=0.904         \pm 0.011$,
$\Cone = 0.143         \pm 0.004$, and
$\Ctwo = -0.326        \pm 0.013$.
Fig.~\ref{fig:Correl}c shows that all finite-size results for the
nematic correlator are nicely described by this fit with the
dependence on $L$ taken into account by substituting Eq.~(\ref{eq:ConfMap})
for the power laws.
Moreover, from $K_+>1$, we see that
the system is indeed in the region dominated by nematic correlations
for $M=3/8$ and $J_1=-J_2$.

Now we turn to the longitudinal correlation function which we fit to the
bosonization result Eq.~(\ref{zzcor}).
Since most numerical parameters have been determined by the previous fit, only
one free parameter is left which we determine from the numerical results
of Fig.~\ref{fig:Correl}a for $L=64$ and $x \ge 14$ as
$\Cthree = 0.060 \pm 0.004$. Predictions for other system sizes are again
obtained by substituting Eq.~(\ref{eq:ConfMap}) for the power laws.
The agreement in Fig.~\ref{fig:Correl}a is
not as good as in Fig.~\ref{fig:Correl}c. However, it improves at larger
distances $x$ and system sizes $L$, indicating that corrections omitted in
Eq.~(\ref{zzcor}) are still relevant on the length scales considered here.

Finally, the $xy$-correlation function is shown in
Fig.~\ref{fig:Correl}b with a  logarithmic scale of the vertical axis of
this panel. The exponential decay predicted by Eq.~(\ref{eq:ssTrans}) is
verified.  One further observes that correlations between the
in-plane spin-operators belonging to different chains (odd $x$)
are an order of magnitude smaller than on the same chain (even $x$).
This suppression of correlations between different chains
corresponds to the $\delta$ symbol in (\ref{eq:ssTrans}),
which strictly applies only in the thermodynamic limit and for
large distances.

We summarize the main result of this section: in-plane spin correlators are exponentially suppressed for any
finite value of the magnetization in the parameter region $|J_1|<J_2$. The ground state crosses over from a 
spin-density-wave dominated to a nematic-like phase
with increasing magnetic field, with the crossover line given by Eq.~(\ref{crossover}).

\section{Excitations}
\label{sec:ex}
We next address the excitation spectrum. Since the gap to $\Delta S^z=1$
excitations should be directly accessible to microscopic experimental probes
such as inelastic neutron scattering or nuclear magnetic resonance,
we  analyze its behavior as a function of magnetization.
Sufficiently below the fully polarized state the gap can be
calculated analytically using results from sine-Gordon theory.
In addition,  to leading order of the interchain coupling, one can get 
qualitative expressions using dimensional arguments
for the perturbed conformally invariant model:
\begin{equation}
\label{Gap}
\Delta_1(m)\sim \left[\frac{c^2(m)|J_1|\sin(\pi m)}{v(m)(1-J_1K(m)/\pi
    v(m))}\right]^{\frac{1}{\nu}} \, ,
\end{equation}
where $\nu=2-2K(m)\big(1+J_1K(m)/\pi v(m)\big)$. 
$m(h), K(h)$ and $v(h)$ can be determined numerically from the
Bethe ansatz integral equations.\cite{Totsuka97,CHP98,Bogoliubov,QFYOA97}

With this information and Eqs.~(\ref{Mag}) and (\ref{Gap})
we determine the qualitative behavior of the single-spin
gap $\Delta_1(M)$ as a function of
$M$: it increases from zero at zero magnetization, reaches a maximum at
intermediate magnetization values, then shows a minimum and, upon
approaching the saturation magnetization, it increases again. 
As our formulas do not strictly apply at $m=0$, the notion of a vanishing gap at zero magnetization
may be a spurious result.
Note that when  the fully polarized state is approached, 
the magnetization increases in an unphysical fashion since in this limit
bosonization becomes inapplicable.
At the point
where the magnetization saturates the exact value of the gap can be
obtained from the following mapping to hard-core bosons:
\cite{Chubukov91,Kuzian,Haldane}
\begin{equation}
S_i^z =\frac{1}{2}-a_i^{\dagger}a_i \, , \qquad
S_i^-= { a_i^{\dagger}} \, . \label{holstein}
\end{equation}
Comparing Eq.~(\ref{holstein}) with Eq.~(\ref{BosFor}) one
recognizes the leading terms in Haldane's harmonic fluid transformation for bosons.\cite{Haldane}
Using a ladder approximation which is exact in the two-magnon subspace  we arrive at:
\begin{eqnarray}
\label{binding}
\Delta_1 \left(M=\frac{1}{2}\right)
&=&
 \frac{4J_2^2-2J_1J_2-J_1^2}{2(J_2-J_1)}
 -\frac{J_1^2+8J_1J_2+16J_2^2}{8J_2}
 \nonumber\\
&=&\frac{1}{8}\,\frac{J_1^2\,(3\,J_2+J_1)}{J_2\,(J_2-J_1)} \, .
\end{eqnarray}
In Eq.~(\ref{binding}) we have
represented the gap as a difference of two terms: the quantum and the 
classical instability fields  emphasizing its {\it quantum origin}.

\begin{figure}[tb]
\includegraphics[width=\columnwidth]{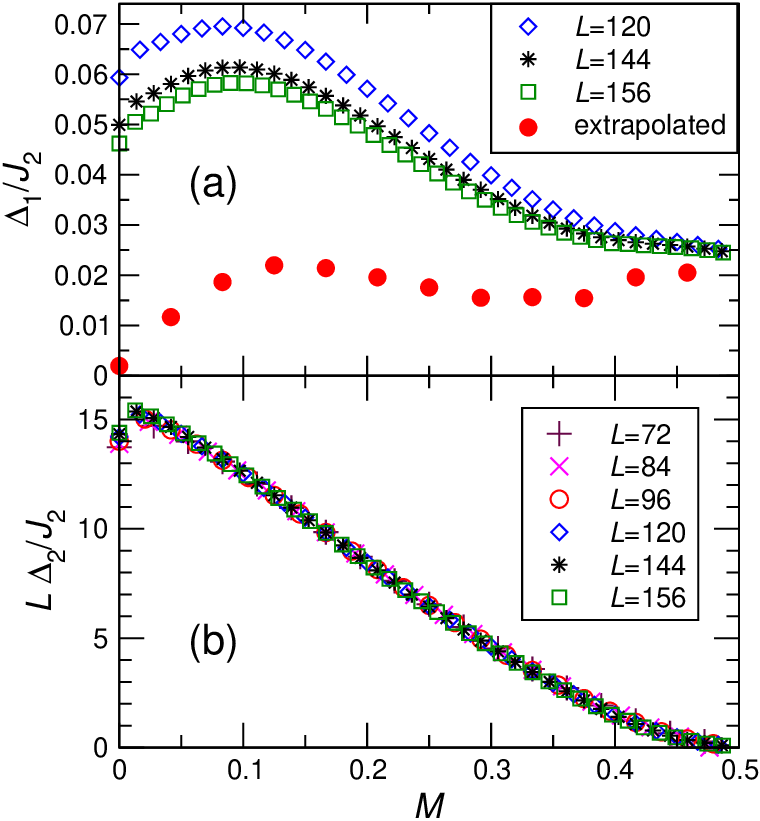}
\caption{(Color online)
Density matrix renormalization group results
for the gaps 
at $J_1 = -0.3\,J_2 < 0$ as a function of magnetization $M$.
Panel (a) shows the single-spin excitation gap (\ref{eq:ssGap}),
panel (b) the finite-size gap (\ref{eq:tsGap}) for two flipped spins
multiplied by the chain length $L$. 
}
\label{fig:gap}
\end{figure}

In order to verify these field theory predictions, we  perform
complementary numerical computations using the
DMRG method. \cite{white92}
Open boundary conditions are imposed and
we typically keep up to 400 DMRG states. From DMRG we
obtain the ground-state energies $E(S^z)$ as a function of total
$S^z$. For those values of $S^z$ that emerge as a ground state
in an external magnetic field we compute the single-spin excitation gap
from
\begin{equation}
\Delta_1(M) = \frac{E(S^z+1)+E(S^z-1)-2\,E(S^z)}{2} \, .
\label{eq:ssGap}
\end{equation} 
Fig.~\ref{fig:gap}a shows numerical results for $\Delta_1$ at a selected
value of $J_1 = -0.3\,J_2 < 0$ for the largest system sizes investigated.
We find that the finite-size behavior of the gap $\Delta_1(M,L)$ for system sizes $L \ge 24$
is well described by a $1/L$ correction. This will be further corroborated
by field-theoretical arguments outlined below. Therefore, we extrapolate
it to the thermodynamic limit using a fit to the form
\begin{equation}
\Delta_1(M,L) = \Delta_1(M) + \frac{a(M)}{L} + \cdots \, ,
\label{eq:extra}  
\end{equation}
allowing for an additional $1/L^2$ correction for those values of
$M$ where at least 4 different system sizes are available.

This extrapolation is represented by the full circles in Fig.~\ref{fig:gap}a;
errors are estimated not to exceed the size of the symbols.
Our extrapolation for
$\Delta_1$ is consistent with a vanishing gap at $M=0$ in agreement with
previous numerical studies \cite{Itoi01} although bosonization
predicts a non-zero -- possibly very small -- gap.
\cite{Itoi01,Nersesyan98,Cabra98}
The behavior of $\Delta_1(M)$ confirms the picture described above:
the gap is non-zero
for $M>0$, goes first through a maximum and then a minimum and finally
approaches  $\Delta_1/J_2 \approx 0.023$ given by
Eq.~(\ref{binding}) for $M \to 1/2$.

We further wish to point out that for chains with periodic boundary
conditions, the coefficient $a(M)$ of the finite-size extrapolation
Eq.~(\ref{eq:extra}) is
determined by the spin-wave velocity and the critical exponent
of the soft mode from the $\Delta S^z=2$ channel. 
Indeed, using Eq.~(\ref{eq:ssTrans}) where we can set $r=0$, and use the conformal mapping (\ref{eq:ConfMap}) to the cylinder, we see
that the leading finite-size correction to  the gap is:
\begin{equation}
\Delta_1(M,L)=\Delta_1(M)+\frac{1}{L}\frac{\pi v_+(M)}{4K_+(M)} \, .
\end{equation}
Note that we have to replace sin with sinh in Eq.~(\ref{eq:ConfMap}) in order
to extract a gap, since we are dealing with Euclidean time.
In addition we used the fact that in our approximation the 
effective Hamiltonian 
(\ref{symantisym}) is a direct sum of symmetric and antisymmetric sectors. 
Moreover, it is only the symmetric sector enjoying conformal invariance and 
consequently we perform the replacement $\tau\to \sinh\tau$ only in
the symmetric sector.
The antisymmetric sector has a spectral gap and its contribution to
the finite-size corrections of the single-spin flip 
excitation energy are exponentially suppressed with system size.\cite{Tsvelick}
With this method one cannot fix the amplitudes of the $1/L^2$ term
and beyond.  Note furthermore that there may be additional surface
terms for open boundary conditions as employed in the numerical DMRG
computations. Nevertheless there is a dominant $1/L$ correction in any case.

Next, we briefly look at the $\Delta S^z=2$ excitations. Their
finite-size gap is, in analogy to Eq.~(\ref{eq:ssGap}), computed with
DMRG from
\begin{equation}
\Delta_{2}(M)=\frac{E(S^z+2)+E(S^z-2)-2\,E(S^z)}{2}\,.
\label{eq:tsGap}
\end{equation}
Fig.~\ref{fig:gap}b shows numerical results for $L\,\Delta_2(M,L)$ again
at the value $J_1 = -0.3\,J_2 < 0$. One observes that the scaled
finite-size gaps collapse onto a single curve which shows that
$\Delta_2(M,L)$ scales linearly to zero with $1/L$, exactly as
expected for gapless excitations in 1D. Furthermore, we observe
that the scaled quantity $L\,\Delta_2(M,L)$ vanishes as one
approaches saturation $M=1/2$ which indicates a vanishing of the velocity
of the corresponding excitations at saturation.

\begin{figure}[tb]
\includegraphics[width=\columnwidth]{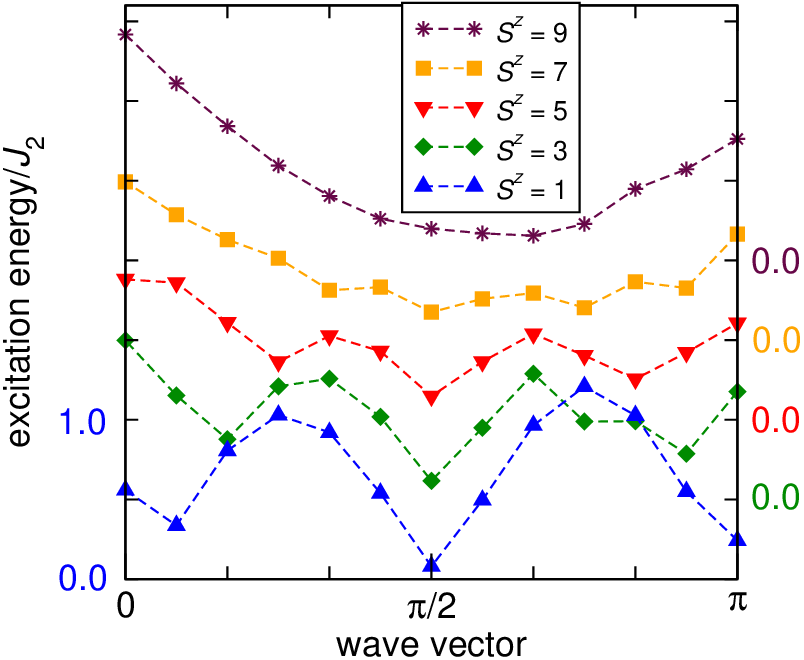}
\caption{(Color online)
Numerical dispersion spectrum in the subspaces of odd $S^z$
computed for $L=24$ and $J_1 = -J_2 <0$. 
The wave vector is given relative to the ground state wave vector (0 for
$S^z =0,4,8$ and $\pi$ for $S^z =2,6$).}
\label{fig:dispersion}
\end{figure}

We proceed by  discussing the wave-vector dependence of the $\Delta S^z =1$
excitation, while we remind the reader that the low-energy excitations are
in the $\Delta S^z=2$ sector.
Fig.~\ref{fig:dispersion} shows representative ED results obtained
for rings with $L=24$ and $J_1 = -J_2 <0$. 
For ground states with low $S^z$, the $\Delta S^z =1$ excitation spectrum
looks similar to the continuum of spinons.
On the other hand, close to saturation one has single-magnon excitations
with a minimum given by the classical value of the wave
vector $k_{cl}=\arccos(|J_1|/4J_2)$.\cite{Chubukov91,Cabra98,GMK98}
We read off from Fig.~\ref{fig:dispersion} that upon
lowering the magnetic field, this
minimum shifts from the classical incommensurate value towards $\pi/2$,
{\it i.e.}\ the value appropriate for two decoupled chains.
This renormalization of the
minimum of the magnon excitations towards the 
value of decoupled chains
can be interpreted in terms of quantum
fluctuations, which are enhanced when the density of magnons increases.
A strong quantum renormalization of the pitch angle from its
classical value at zero magnetization was previously observed by
the coupled-cluster method and DMRG calculations.\cite{Bursill}

\section{Summary}
\label{sec:sum}
We have combined numerical techniques with analytical approaches and mapped out the ground state phase diagram of the frustrated ferromagnetic spin chain
in an external magnetic field. We have established that with increasing magnetic field, the ground state crosses over
from a spin-density-wave dominated 
to a nematic-like phase.
Single spin flip excitations are gapped, giving rise to an exponential decay
of in-plane spin correlation functions in  both  regimes.
We have studied the single- and two-spin flip excitation energy numerically. 
Using tools from conformal field theory we have further shown that the amplitude of the leading $1/L$ correction term to the
single-spin flip gap is determined by the critical exponent and the spin-wave velocity of the soft mode.  

Finally, in order to apply our findings to the material LiCuVO$_4$, one should take into account 
interchain interactions as well as anisotropies, which are expected to be present in 
this system.\cite{Enderle05}
At low fields, a helical state has been observed
experimentally.\cite{Gibson04,Enderle05}
On the other hand, for the purely one dimensional case, we have shown
that upon increasing the magnetic field there is a competition between
spin-density-wave and nematic-like tendencies. Those are the 
two leading instabilities at high magnetizations and thus they are the natural candidates to become long-range 
ordered in higher dimensions.
The question whether there are true phase transitions at high fields in higher dimensions is beyond the scope of the current work.

\acknowledgments

We thank A.~Feiguin for  providing us with his DMRG
code used for large scale calculations.  Most  of T.V.'s work was done
during his visits to the 
Institutes of Theoretical Physics at the Universities
of Hannover and G\"ottingen, supported by the Deutsche Forschungsgemeinschaft. 
The hospitality of the host institutions is gratefully acknowledged.
T.V.\ also acknowledges support from the Georgian National Science Foundation under grant 
N~$06_-81_-4-100$. LPTMS is a mixed research unit 8626 of CNRS and University Paris-Sud.
A.H.\ is supported by the Deutsche Forschungsgemeinschaft (Project No. HO~2325/4-1), and
F.H.-M.\ is supported by NSF grant No. DMR-0443144.

\end{document}